\documentclass[aps,prd,nofootinbib]{revtex4}
\usepackage[colorlinks=true, pdfstartview=FitV, linkcolor=blue, citecolor=red, urlcolor=magenta]{hyperref}
\usepackage{graphicx}
\usepackage{latexsym}
\usepackage{amsmath}
\usepackage{amsfonts}
\usepackage{amssymb}
\usepackage{verbatim}

\newcommand{\be}{\begin{equation}}
\newcommand{\ee}{\end{equation}}
\newcommand{\bea}{\begin{eqnarray}}
\newcommand{\eea}{\end{eqnarray}}

\newcommand{\pa}{\partial}
\newcommand{\bb}{\bibitem}

\def\bb{\bibitem}

\def\bb{\bibitem}
\newcommand{\ben}{\begin{eqnarray}}
\newcommand{\een}{\end{eqnarray}}

\newcommand{\remark}[1]{}

\begin{document}
\title{Lorentz-violating dimension-five operator contribution to the black body radiation}

%\title{Corrections to black-body radiation spectrum with Lorentz-violating dimension-five operator}

%%%%%%%%%%%%%%%%%%%%%%%%%%%%%%%%%%%%%%%%%%%%%%%%%%%%%%%%%%%%%%%

\author{$^{1}$M. A. Anacleto}
\email{anacleto@df.ufcg.edu.br}
\author{ $^{1,2}$F. A. Brito}
\email{fabrito@df.ufcg.edu.br}
\author{$^{1,2}$ E. Maciel}
\email{eugeniobmaciel@gmail.com}
\author{$^{3}$ A. Mohammadi}
\email{azadeh.mohammadi@df.ufpe.br}
\author{$^{1}$E. Passos}
\email{passos@df.ufcg.edu.br}
\author{$^{1,2}$ W. O. Santos}
\email{wagner$_08$@hotmail.com}
\author{$^{1}$ J. R. L. Santos}
\email{joaorafael@df.ufcg.edu.br}
%%%%%%%%%%%%%%%%%%%%%%%%%%%%%%%%%%%%%%%%%%%%%%%%%%%%%%%%%%%%%%%
 
\affiliation{$^{1}$Departamento de F\'{\i}sica, Universidade Federal de Campina Grande,\\
	Caixa Postal 10071, 58429-900, Campina Grande, Para\'{\i}ba, Brazil.}
\affiliation{$^{2}$Departamento de F\' isica, Universidade Federal da Para\' iba,\\  Caixa Postal 5008, Jo\~ ao Pessoa, Para\' iba, Brazil.}

\affiliation{$^{3}$Departamento de F\' isica, Universidade Federal de Parnambuco,\\  Caixa Postal 171-900, Recife, Pernambuco, Brazil.}

%\preprint{}

\begin{abstract}
We investigate the thermodynamics of a photon gas in an effective field theory model that describes Lorentz violations through dimension-five operators and Horava-Lifshitz theory. We explore the electrodynamics of the model which includes higher order derivatives in the Lagrangian that can modify the dispersion relation for the propagation of the photons. We shall focus on the deformed black body radiation spectrum and modified Stefan-Boltzmann law to address the allowed bounds on the Lorentz-violating parameter.
\end{abstract}
\pacs{XX.XX, YY.YY} \maketitle

%\vspace{1cm}

\section{Introduction}

The Lorentz symmetry has been tested in many experiments at high energy. The possibility of being an exact or approximate symmetry has been addressed in many studies in the literature. Most of these studies are in essence inspired  by Lorentz-violating Standard Model Extension proposed by Colladay and Kosteleck\' y \cite{ck}. However, we shall first focus on thermal dynamics from the theories with higher derivative Lorentz-violating operators of quantum electrodynamics extended with dimension-five operators proposed by Myers and Pospelov \cite{mp} and then extend the analyses to the Horava-Lifshitz theory \cite{horava}. One of the main idea concerning such an issue is to assume that the Lorentz symmetry can be broken at very high energy. 
Recently, higher derivative or higher dimensional operators have been used in quantum gravity \cite{horava} and field theory \cite{mp,km,tm,lopez,ca} to explore new physics in the Lorentz symmetry violation context.

%2) falar de running ou flow de Lorentz-violating parameters como em Horava theory (PRL)

In more recent studies the issue of higher derivative theories has been considered in the context of Horava-Lifshitz (HL) gravity \cite{horava} which aims to solve the severe problems of quantum gravity. In such setup the theory breaks the Lorentz symmetry in the ultraviolet (UV) regime and recovers the symmetry at low energy --- infrared (IR) regime. 
In this theory, there exists an explicit asymmetry between time and space coordinates such that the Lorentz symmetry is broken.
The asymmetry flows between UV and IR scales according to a critical exponent that is also known as Lifshitz critical exponent \cite{HL-PRL}. This dynamical asymmetry makes the HL gravity a nonrelativistic theory that develops an anisotropic scaling
symmetry of space and time. The higher spatial derivative terms added to the action enforces the theory power counting renormalizable.

%{\bf 1) falar mais de operadores 5d (Myers-Pospelov)}
%As we shall see later, the construction of these operators is inspired in the essence of Lorentz-violating Standard Model Extension proposed by Colladay and Kosteleck\' y \cite{ck}. 

%However, we shall focus on thermal dynamics from the theories with higher derivative Lorentz-violating operators of quantum electrodynamics extended with dimension-five operators proposed by Myers and Pospelov \cite{mp}. 
In this work we shall both consider the gauge sector of Myers-Pospelov and Horava-Lifshitz theory  in order to study the modifications in the thermodynamic properties of a photon gas in such a scenario where we have Lorentz-violating operators. 
It is natural to  expect that fingerprints of Lorentz-violation will appear in observations involving cosmic microwave
background (CMB) and ultra high energy sources such as cosmic rays or gamma ray bursts (GRB). 
As is well-known the CMB radiation develops a black body spectrum very accurate for low frequencies. 
However one expects that for more accurate measurements deviations from black body radiation at high frequencies may become the place to find the signs of Lorentz-symmetry violation.
It is remarkable that fingerprints from such operators can be seen in the deformed black body radiation when $z$ is around (above or below) unit at CMB temperatures.  Furthermore, whereas stronger modification for lower dimensional operators is favorable at CMB scale, stronger contribution from higher dimensional operators takes places at larger scales.
%{\bf 3) falar da radia‹o de corpo e suas modfica›es em geral (Balachandra-Amilcar, Brito-Barosi-Amilcar-JCAP, Brito-Elisama-IJMPA, etc)

There are some previous investigations on the modification of the black body radiation due to Lorentz-violating theories --- see eg. \cite{casana2008} and particularly \cite{Balachandran:2007ua} for a noncommutative field theory.  In the latter case, such theories present a deformed black body spectrum whose deviations from the usual one appear in the ultraviolet regime. This effect has been considered to find new bounds on the noncommutative parameters of the theory through the, for instance, GZK cut-off. In addition this modification at UV is in direct connection with the aforementioned theories whose Lorentz-symmetry violation arises due to influence of higher derivative terms. In Ref.~\cite{bbq2007} was found indeed that 
at the temperature around $T = 10^{14}$GeV $\simeq10^{27}$K, which approximately corresponds to the temperature at the beginning of the inflation, the noncommutative parameters are in agreement with the bounds obtained through the GZK cut-off  which is of the order of $10^{19}$eV (the highest cosmic ray energy is assumed to be no greater than $10^{19}$eV) More investigations on the deformed thermodynamics in noncommutative theories can be found in Ref.~\cite{Brito:2015csa} where new bounds to the noncommutative parameters can also be found by using the critical temperature of a deformed Bose-Einstein condensate. In the following we delineate our investigations by following these ideas to address similar issues to the deformed black body in the present Lorentz-violating theory due to dimension-five operators.

%}

%%%%%%%%%%%%%%%%%%%%%%%%%%%%%%%%%%%%%%%%%%%%%%%%%%%%%%%%%%%%%%%%%%%%%%%%%%%%%%%%%
The outline of this paper is as follows. In Sec.~ \ref{II}, we introduce the Myers-Pospelov model of electrodynamics
with a Lorentz-violating background. The  dispersion relation is obtained in a preferred frame defined by a time-like direction. In Sec.~\ref{SBL}, we address the issues of the modified Stefan-Boltzmann law. In Sec.~\ref{HL-temperature} we extend the previous analysis to Horava-Lifshitz theory and in Sec.~\ref{conclu} we make our final considerations.

%%%%%%%%%%%%%%%%%%%%%%%%%%%%%%%%%%%%%%%%%%%%%
\section{The dimension-five operator and Thermodynamic Properties}

\subsection{Modified Dispersion Relation}
\label{II}
%\section{Photons with dimension-five operator}
%In this section, we derive the dispersion relations associated with the effective Lagrangian Maxwell
%term supplemented by dimension-five operators given in the following gauge invariant Lagrangian:
%\bea\label{smg01}
%{\cal L}=-\frac{1}{4}F_{\mu\nu}F^{\mu\nu} - g\, \varepsilon^{\alpha\mu\lambda\rho}n_{\alpha} n^{\nu} n^{\sigma} (\pa_{\lambda}F_{\mu\nu} )F_{\rho\sigma}.
%\eea
%where $g = \bar{\xi}/M$, with $\bar{\xi}$ being a dimensionless parameter. The dimension-5
%operator in this electromagnetic sector is CPT - odd and charge conjugation - even, and $M$ is the mass where new physics such as Lorentz and CPT symmetry violation emerges. 

%Thus, the new equation of motion for the vector potential $A_{\mu}=(A_{0},A_{i})$ is given as
%\bea\label{m6}
%\big(\pa^{2}\eta^{\mu\nu}-\pa^{\mu}\pa^{\nu}-2g(n\cdot\pa)^{2}\varepsilon^{\mu\nu\lambda\rho}n_{\lambda}\pa_{\rho}\big)A_{\nu}=0.
%\eea
As mentioned above, the Lorentz-violating dimension-five operator predicts a modified dispersion relation \cite{mp}, which can be
characterized by the following covariant form \cite{Reyes}:
\bea\label{m7}
k^{4}-4\,{\bar{\xi}}^{2}(n\cdot k)^{4}\Big[(k\cdot n)^{2}-k^{2}n^{2}\Big]=0\,.
\eea
%In this point we also make a simple analysis of solutions to the dispersion, Eq.(\ref{m7}),
Such a dispersion relation reduces to its simplest form when the parameter $n_{\mu}$
is chosen to be purely timelike, i.e., $n_{\mu}\equiv(1,\vec{0})$: 
\bea\label{m8}
\bold k^{2}=  E^{2} \big(1 - 2\lambda \bar{\xi} E\big)^{-1}\,,
\eea
where in this last form we work with the notation: $k^2=k_{\mu} k^{\mu}=(E^2, -\bold k^2)$, $\bold k^2=k_i k_i$ and $i=1,2,3$. We should stress that the parameter $\lambda$ characterizes two polarization states, which are explicitly given by $\lambda=\pm 1$. Moreover, the standard dispersion relation can be recovered by setting $\bar{\xi}\to 0$. Notice also that the phase and group velocities derived from Eq.~(\ref{m8}) are related through Rayleigh's formula \cite{epassos_17}. By considering a  subluminal case $\lambda=-1$ we consequently have that $v_{p} > v_{g}$, characterizing a normal dispersion medium. Moreover,  in the superluminal case we take $\lambda=1$ resulting in $v_{g} > v_{p}$, or in another words, we have an anomalous medium which is related to anisotropic effects. Therefore, we can conclude that a model truly isotropic must be attributed only to subluminal case. So, in our thermodynamic analyses, we consider only the normal dispersion medium corresponding to $\lambda=-1$.

The above configuration corresponds to a subset of Lorentz invariance violating (LIV) operators which preserves the rotational invariance. Such an isotropic inertial frame must be specified, once boosts to other frames can destroy the rotational invariance. One natural choice for the preferred frame is the frame of the Cosmic Microwave Background (CMB), however other frames are also available, as it was discussed in \cite{Kost_Mewes_01}. 

In this work, we consider astrophysical limits to $\bar{\xi}$ and use them to analyze the black body radiation due the modified dispersion relation given by Eq.(\ref{m8}). The bounds for these astrophysical limits were derived from collected data \cite{kost_russ} (see the first and the second lines in the table \ref{TAB0}), and from simultaneous gravitational and electromagnetic waves \cite{epassos_17} (see the third line in the table \ref{TAB0}).

\begin{table}[h!]
\centering
	\begin{tabular}{|c|c|c|c|} 
		\hline
		{\;\; Adjustment\;\; }&{\;\;Result ($\rm GeV^{-1} $)} &{\;\; Result (degree kelvin - ${\rm k}^{-1}$)}\;\; &{\;\; System}\;\;  \\ \hline
		$\bar{\xi}\equiv k^{(5)}_{(V)00}$  &\;\;\; $\sim 10^{-32}$ & \;\; $\sim 10^{-45}$ &{\;\; Astrophysical
birefringence }\;\;  \\ \hline
			$\bar{\xi}\equiv k^{(5)}_{(V)00}$ & \;\;\; $\sim 10^{-20}$  &  $\;\;\sim 10^{-33}$ &{\;\; CMB polarization}\;\;    \\  \hline
			\!\!\!\!\!\!\!\!\!\!$\bar{\xi}\equiv \bar{\xi}_{\gamma}$ &\;\;\;  $\;\sim 10^{-13}$   &   \;\; $\sim 10^{-26}$  &{\;\; GW-GRB waves}\;\;  \\  \hline	
	\end{tabular}
	\caption{Astrophysical bounds for the Lorentz-breaking parameter.}
	\label{TAB0}
\end{table}

%%%%%%%%%%%%%%%%%%%%%%%%%%%%%%%%%%%%%%%%%%%%%%%%%%%%%%%%%%%%%%%%%%%%%%
\subsection{The partition function}
In order to study the thermodynamical features of Myers-Pospelov model, we need to build its partition function. As it is known, the first ingredient to derive the partition function is the establishment of the number of states. In general, the number of available states for a given system can be written as
\bea\label{T01a}
\Omega = \frac{\gamma}{(2 \pi)^{3}} \int\int d\vec{r}d\vec{k}\,,
\eea
with $\gamma$ being the spin multiplicity ($\gamma=2$ for photons). The last equation can be rewritten in spherical coordinates as
\bea\label{T01}
\Omega=\frac{V}{\pi^{2}}\int_{0}^{\infty}{\bold k}^{2}d{\bold k}\,,
\eea
where $V$ is the volume of the reservoir.
Then we can substitute ${\bold k}^{\,2}$ by the dispersion relation (\ref{m8}), leading us to
\bea\label{T02a}
d{\bold k} =\left( \frac{1}{\sqrt{1+2\, E\,  \bar{\xi} }}-\frac{E\,  \bar{\xi} }{(1+2\,E\,  \bar{\xi} )^{3/2}}\,\right) dE.
\eea
%\bea\label{T02}
%k^{2}dk= E^{2}(1 + 4 g E %+\frac{15 E^2 g^2}{2}
%+\mathcal{O}[g])dE
%\eea
Therefore, the number of states can be represented in the following form
\bea\label{T03}
\Omega =\frac{V}{\pi^{2}}\int_{0}^{\infty} E^2\,(1+E\, \bar{\xi})\,\left(1+2 \,E \,\bar{\xi}\right)^{-5/2}\,dE\,.
%(1 + 4 g E+\frac{15 E^2 g^2}{2}+\mathcal{O}[g])
\eea
%Notice that the LIV effects change the number of states available to the system, increasing as $ \lambda = - 1 $ and decreasing as $ \lambda = + 1 $. 

Let us now proceed to calculate the thermodynamic properties of such a
photon gas. In order to achieve this objective,  let us remember that the connection between
macroscopic world and the thermodynamic behavior is done by the partition
function ${\cal Z}$, i.e., the knowledge of this function allows us to deduce all the
thermodynamics of the corresponding system. The partition function can be derived from
\bea\label{T04}
\ln {\cal Z}(\beta, V) =-\frac{V}{\pi^{2}}\int_{0}^{\infty} E^2\,(1+E\, \bar{\xi})\, \left(1+2 \,E \,\bar{\xi}\right)^{-5/2}\,\ln\big[1 - e^{-\beta E}\big]dE\,.
\eea
This equation is the familiar expression for Bose-Einstein statistics modified by LIV terms \cite{casana2008}. Here  $\beta =T^{\,-1}$ (for $k_B=1$ in natural units), where $T$ is the temperature of the black body, which means, the temperature of the Universe, in our approach.

%%%%%%%%%%%%%%%%%%%%%%%%%%%%%%%%%%%%%%%%%%%%%%%%%%%%%%%%%%%%%%%
\subsection{Correction to Stefan-Boltzmann law}
\label{SBL}

\begin{figure}[h!]
 \centering
 \includegraphics[width=0.4\columnwidth]{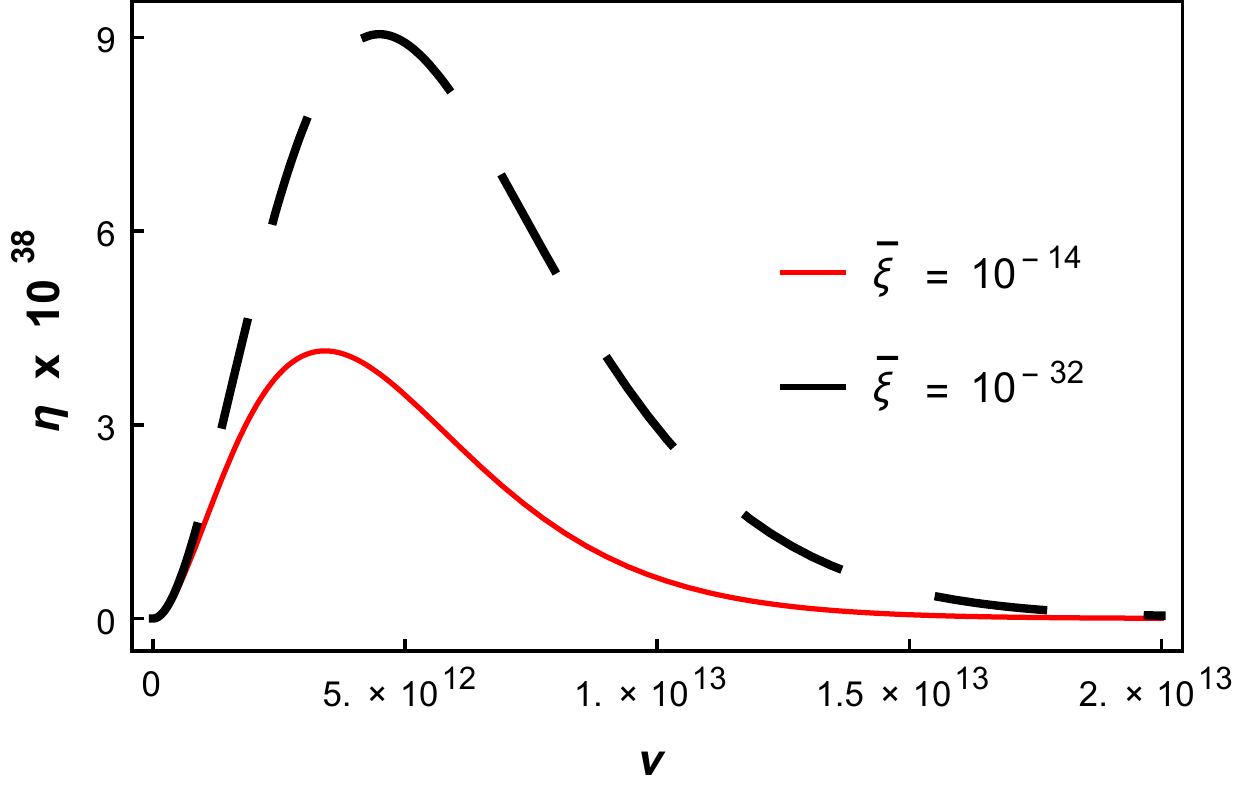} \hspace{0.2 cm} \includegraphics[width=0.4\columnwidth]{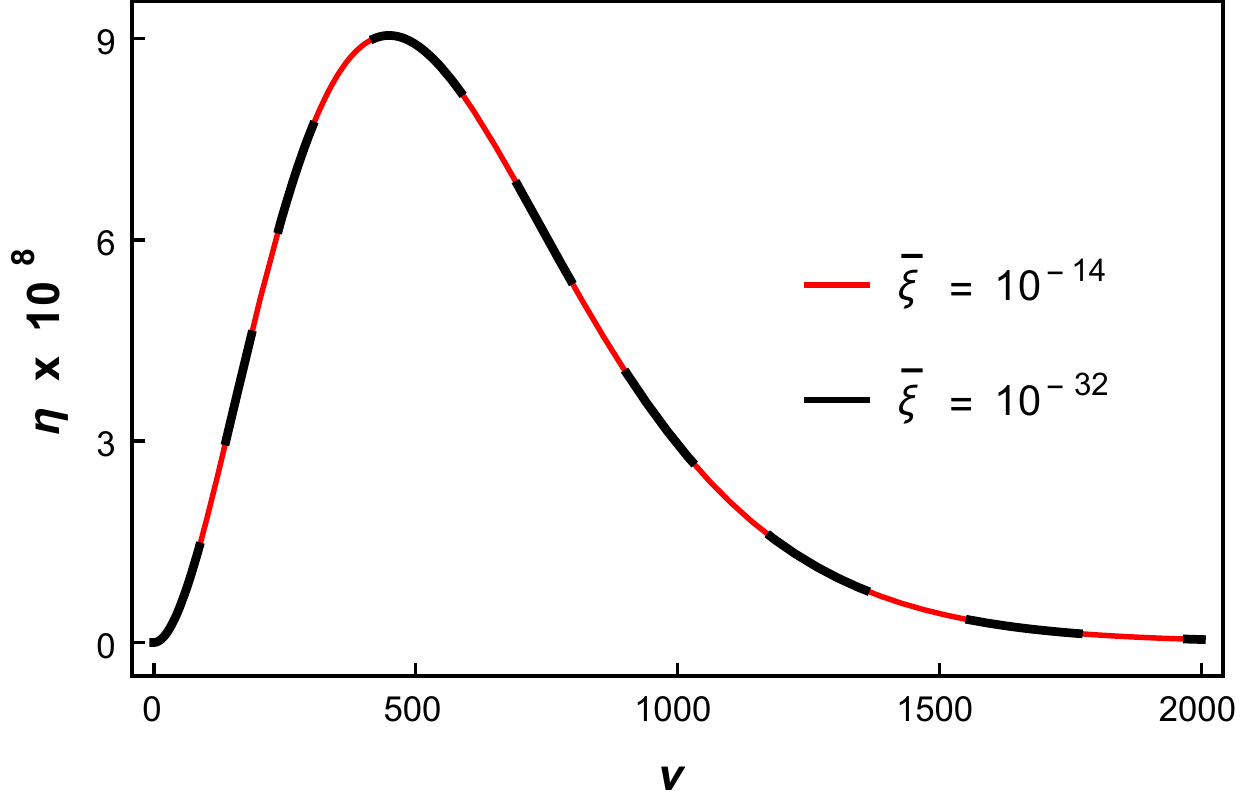}
 \caption{Black body radiation curves as a function of the frequency $\nu$. In the left panel we see the spectral radiance for the inflationary era of the Universe ($T=10^{\,13}$ GeV). Besides, the right panel shows the spectral radiance for the electroweak epoch of the Universe ($T=10^{\,3}$ GeV). In both graphs the units for $\eta$ and $\bar{\xi}$ are GeV$^{\,3}$ and GeV$^{\,-1}$, respectively.}
 \label{FIG1}
\end{figure}

The study of the Stefan-Boltzmann law associated with a system can be done by determining its internal energy density. The internal energy of this Lorentz-breaking model has the form 
\bea\label{T05}
U=-\left(\frac{\pa \ln {\cal Z}(\beta, V)}{\pa \beta}\right)_{V}\,,
%\frac{U}{V}
%\nonumber\\p&=&\frac{1}{\beta V}\ln {\cal Z}(\beta, V).
\eea
which can be used to derive the spectral radiance $\eta(\nu)$. The spectral radiance is explicitly given by
\be
\eta=\frac{E^3 \left(1+E \,  \bar{\xi}\right) }{\pi ^2 \left(e^{\beta\, E}-1\right)} \, (1+2 E \, \bar{\xi} )^{-5/2}\,;\qquad E = h\,\nu\,,
\ee
whose behavior is plotted in Fig.~\ref{FIG1}. There we can see that the Lorentz-breaking parameter is truly effective only if we deal with high temperatures for the Universe, such as the inflationary epoch temperature, when $T=10^{\,13}$ GeV.

\begin{figure}[h!]
 \centering
 \includegraphics[width=0.45\columnwidth]{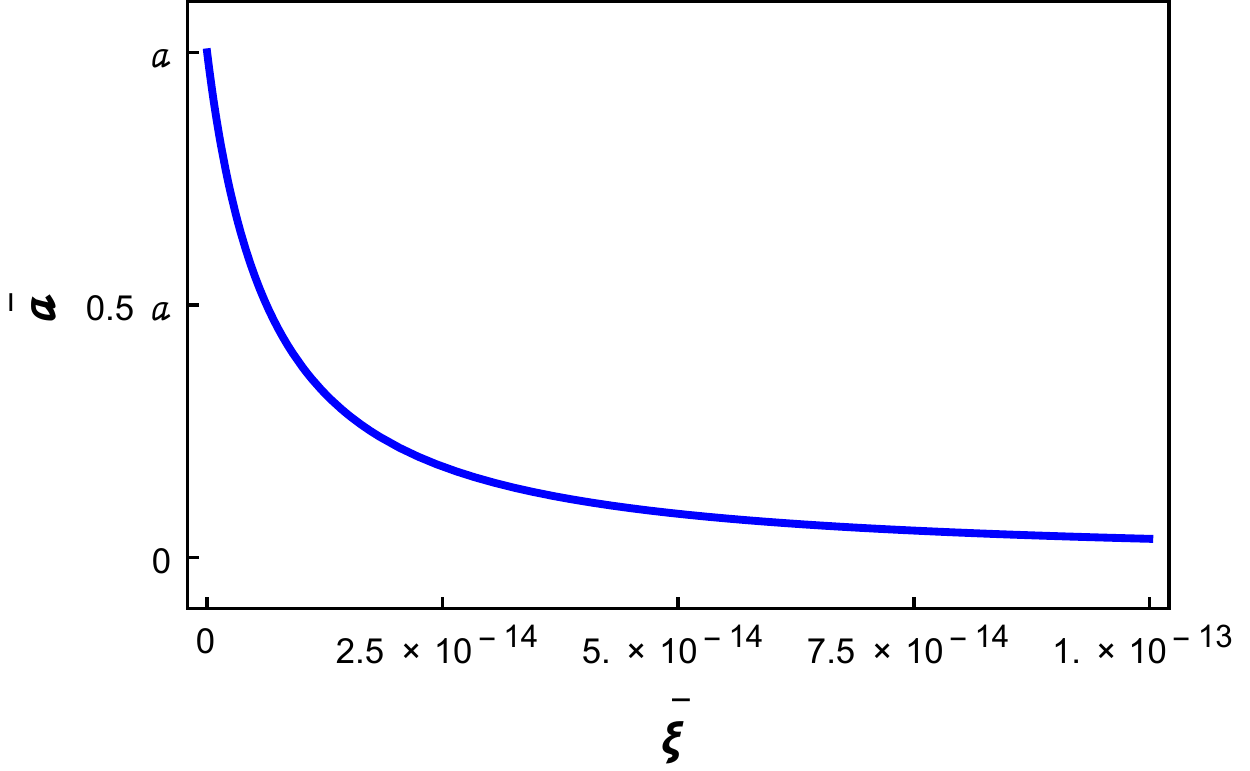} \hspace{0.2 cm} \includegraphics[width=0.45\columnwidth]{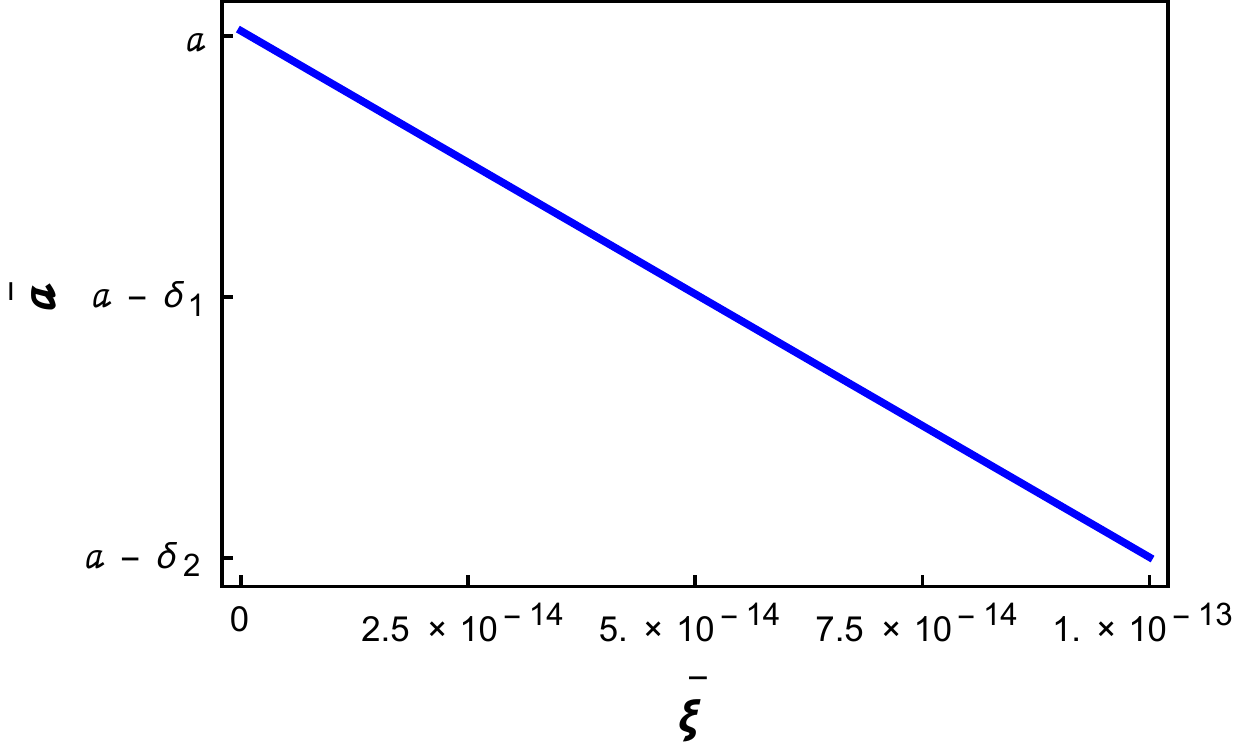}
 \vspace{0.5 cm}
 \includegraphics[width=0.45\columnwidth]{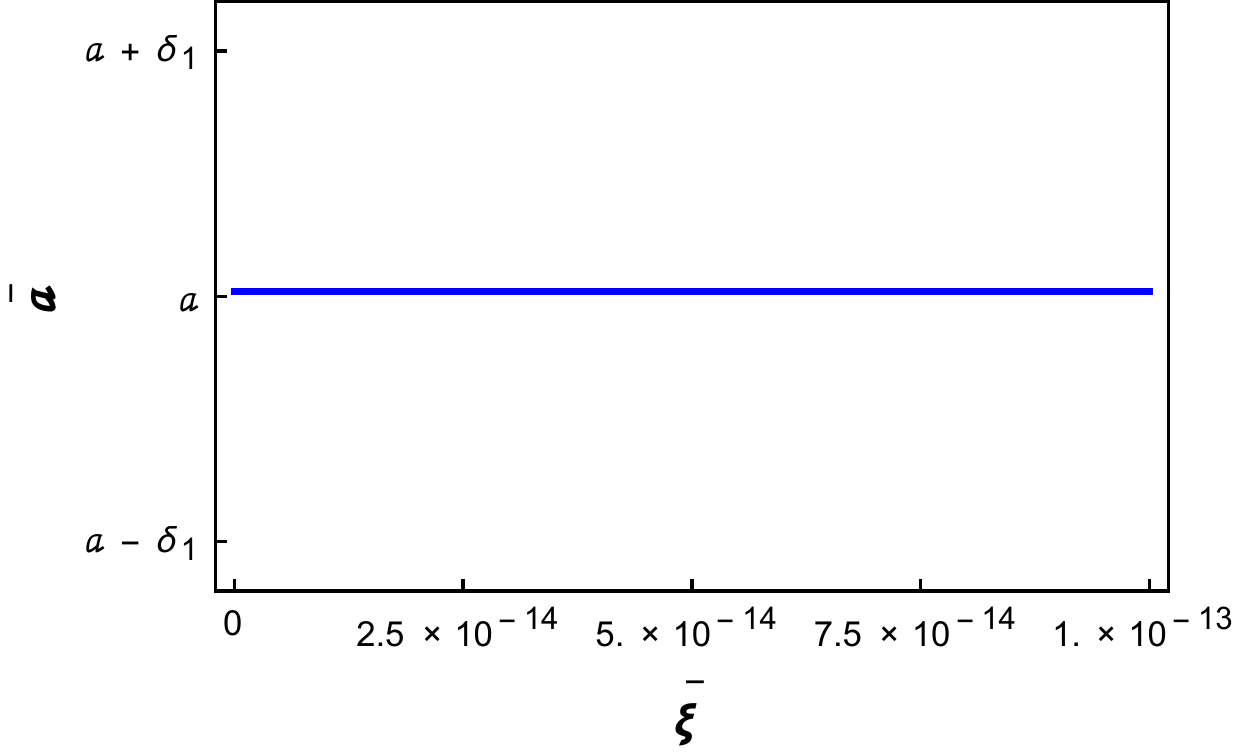}
 \caption{Modified radiation constant $\bar{a}$ versus the Lorentz-breaking parameter for different sets of temperatures. The upper graph in the left shows the behavior of $\bar{a}$ at the inflationary era ($T=10^{\,13}$ GeV). Besides, the upper right panel unveils the evolution of the modified radiation constant at the electroweak epoch of the Universe ($T=10^{\,3}$ GeV), there, $\delta_1=5.0\,\times\,10^{\,-10}$ GeV while $\delta_2=10^{\,-9}$ GeV. Moreover, in the lower graph we can see that $\bar{a}=a$ at the CMB temperature ($T=2.35\times 10^{-13}$ GeV).}
 \label{FIG2}
\end{figure}

Moreover, Eq. $(\ref{T05})$ allows us to derive
\bea\label{BB01}
u(\beta, \bar{\xi}) = \frac{U}{V}= u_{_{1}}(\beta, \bar{\xi}) + u_{_{2}}(\beta, \bar{\xi})
\eea
which is the internal energy density for the Myers-Pospelov black body. In this last equation, we have
\begin{subequations}
\bea\label{BB02}
u_{_{1}}(\beta, \bar{\xi})= \frac{1}{\pi^{2}}\int_{0}^{\infty}\frac{E^{3}\big(1 + 2 \bar{\xi} E\big)^{-5/2} dE}{e^{\beta E} -1}, 
\eea
\bea\label{BB03}
u_{_{2}}(\beta, \bar{\xi})= \frac{\bar{\xi}}{\pi^{2}}\int_{0}^{\infty}\frac{E^{4}\big(1 + 2 \bar{\xi} E\big)^{-5/2} dE}{e^{\beta E} -1}. 
\eea
\end{subequations}
The desired behavior for $u$ in the limit $\bar{\xi}\rightarrow 0$ or if we choose low temperatures scenarios, is to find $u=u_{s}=a\,T^{\,4}$, where $u_{s}$ is the Stefan-Boltzmann energy density, and $a$ is the so-called radiation constant ($a=4\,\sigma=\pi^{\,2}/15$ in natural units). In order to see how the Lorentz-breaking parameter can affect the radiation constant, let us rewrite $(\ref{BB01})$ in the following way
\be \label{BB04}
\bar{a}\equiv u\,\beta^{\,4}= \left( u_{_{1}}(\beta, \bar{\xi}) + u_{_{2}}(\beta, \bar{\xi})\right)\,\beta^{\,4}.
\ee
%Unfortunately, we are not able to solve the last equation analytically, however, 
We can analyze the evolution of $\bar{a}$ via a numerical integration. By recalling that $E=h\,\nu$, we can perform a numerical integration of $(\ref{BB04})$ with respect to the frequency. Such a procedure leads us to the graphs plotted in Fig.~\ref{FIG2}. We depicted three graphs in this figure, with the Lorentz-breaking parameter $\bar{\xi}$ variating from $10^{\,-32}$ up to $10^{\,-13}$. There, we can see that when the Universe was really hot, such as slightly before its primordial inflationary phase, the radiation constant was really different from its actual value (upper panel in the left). The upper graph in the right side of Fig.~\ref{FIG2} describes how $\bar{a}$ approaches to $a$, when the Universe enters in the electroweak era. Moreover, the lower panel unveils that at the CMB (Cosmic-Microwave-Background) temperature ($T=2.725$ K $=$ $2.35\,10^{\,-13}$ GeV), the radiation constant is exactly equal to $\pi^{\,2}/15$ for $\bar{\xi}$ presented in Table \ref{TAB0}.

\begin{figure}[h!]
 \centering
 \includegraphics[width=0.45\columnwidth]{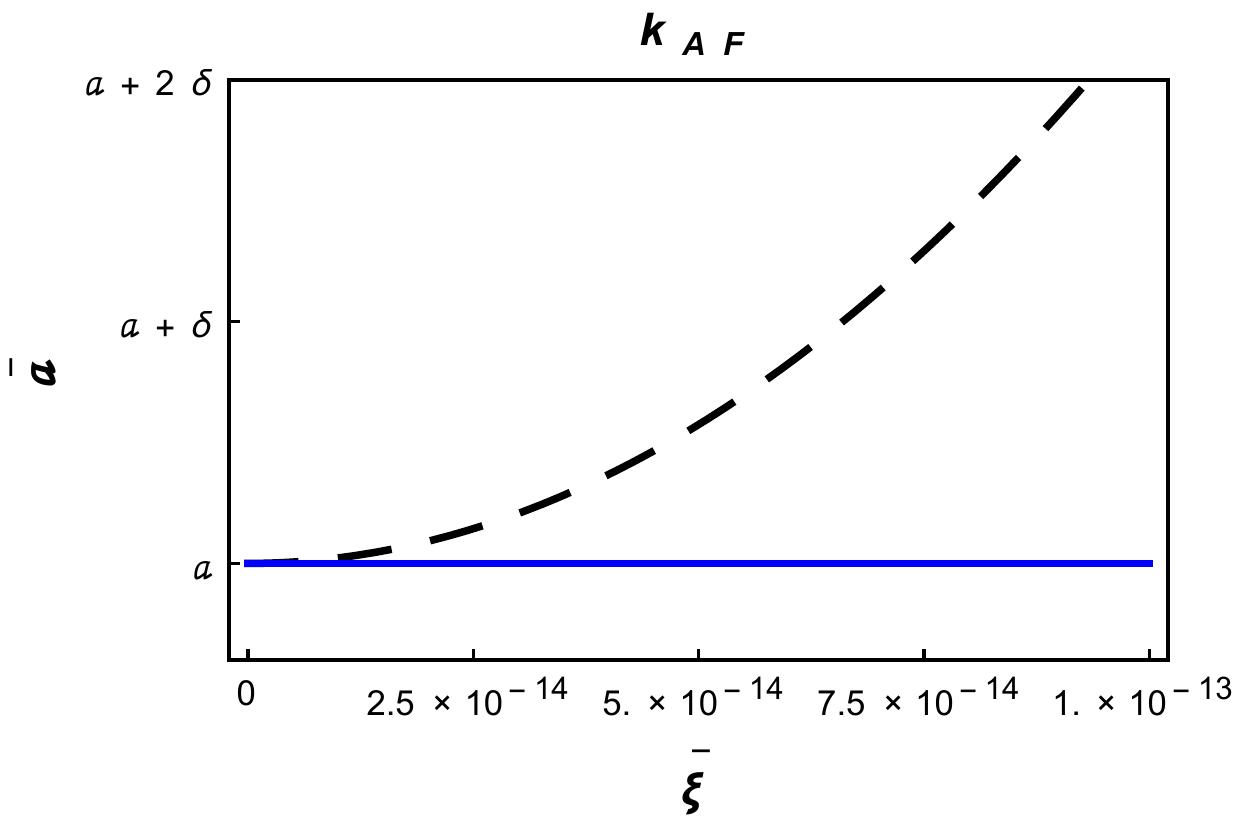}
 \caption{Modified radiation constant $\bar{a}$ versus the Lorentz-breaking parameter $\bar{\xi}$ and versus $k_{\,AF}$ at the CMB temperature ($T=2.35\times 10^{-13}$ GeV), where $\delta=2.5\,\times\,10^{\,-3}\,a$.}
 \label{FIG3}
\end{figure}

Another interesting feature of our modified radiation constant $\bar{a}$, emerges when we compare it with the one found in \cite{casana2008}. There, the authors determined a modified Stefan-Boltzmann law, by analyzing the thermodynamic properties from a Maxwell Lagrangian coupled to the Carroll-Field-Jackiw term  \cite{cfj1990}. In their approach, they yield to  the relation
\be
u_{MCFJ}=\bar{a}\,T^{\,4}\,;\qquad \bar{a}=a+\frac{k_{AF}^{\,2}}{48\,T^{\,2}}\,,
\ee
where $k_{AF}$ is their Lorentz-breaking parameter. We also point that in their work, they constrained such a parameter using the CMB temperature. The evolution of $\bar{a}$ at the CMB temperature, with respect to both $\bar{\xi}$ and $k_{AF}$ is depicted in Fig.~\ref{FIG3}. There we can see that $k_{AF}$ is sensible to this scale of temperature, while $\bar{\xi}$ is not. This behavior is based on the fact that the Myers-Pospelov Lagrangian is written in terms of a dimension-five operator, which is higher than the one coming from the Maxwell-Carroll-Field-Jackiw Lagrangian. So, as predictable, a higher dimension operator is going to be probed only at higher energy scales, unveiling the consistence of our thermodynamical approach. Furthermore, as we shall show below, one can flow between higher to lower derivative operators --- compared to dimension-five operator. One precisely finds that as higher derivative operators take place they lose their contributions to lower derivative operators at CMB scale.

%%%%%%%%%%%%%%%%%%%%%%%%%%%%%%%%%%%%%%%%%%%%%%%%%%%%%%%%%%%%%%%%%%%%%%%%%%%%%%%%%%%%%%%%
%%%%%%%%%%%%%%%%%%%%%%%%%% - FIGURES - %%%%%%%%%%%%%%%%%%%%%%%%%%%%%%%%%%%%%%%%%%%%%%%%%

%%%%%%%%%%%%%%%%%%%%%%%%%%%%%%%%%%%%%%%%%%%%%%%%%%%%%%%%%%%%%%%%%%%%%%%%%%%%%%%%%%%%%%%%
\section{The Lifshitz-scaling and Thermodynamic Properties}
\label{HL-temperature}
\begin{figure}[h!]
 \centering
 \includegraphics[width=0.4\columnwidth]{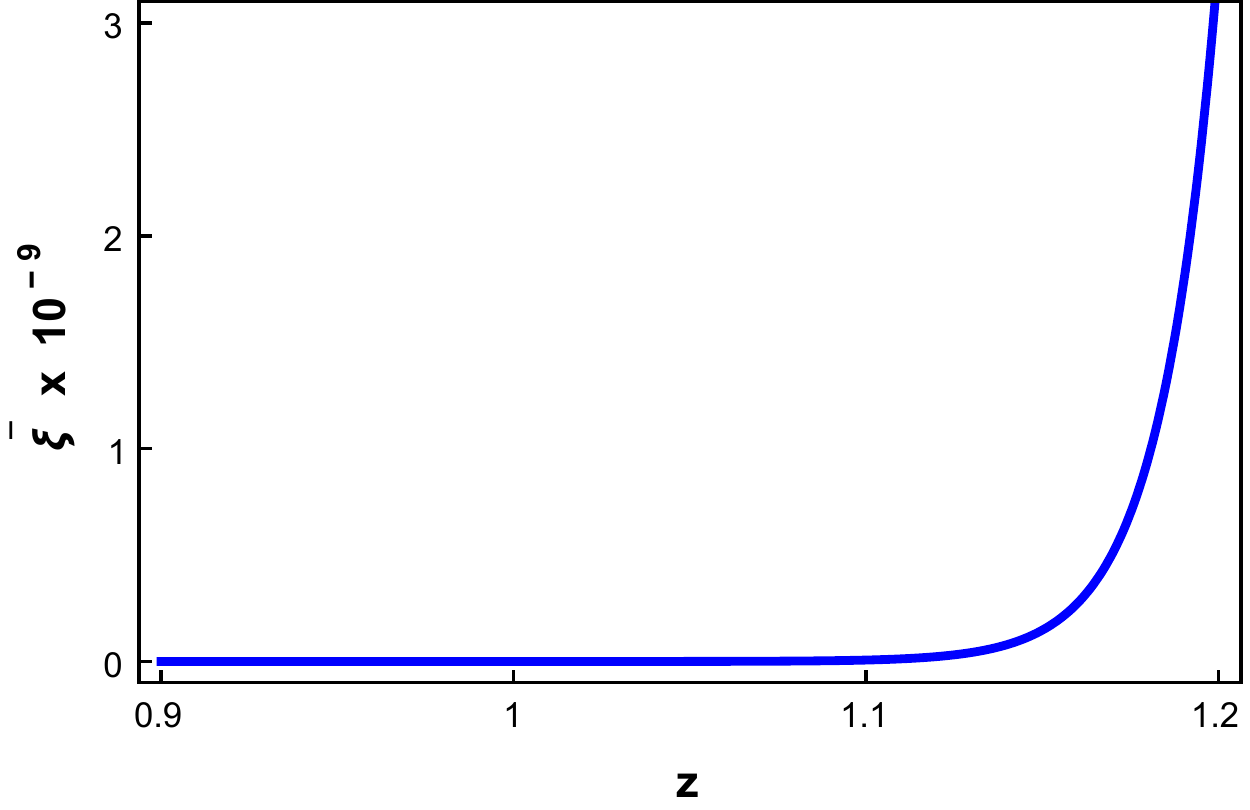}
 \caption{Lorentz breaking parameter as function of the Lifshitz critical exponent $z$.}
 \label{FIG4}
\end{figure}

\begin{figure}[h!]
 \centering
 \includegraphics[width=0.4\columnwidth]{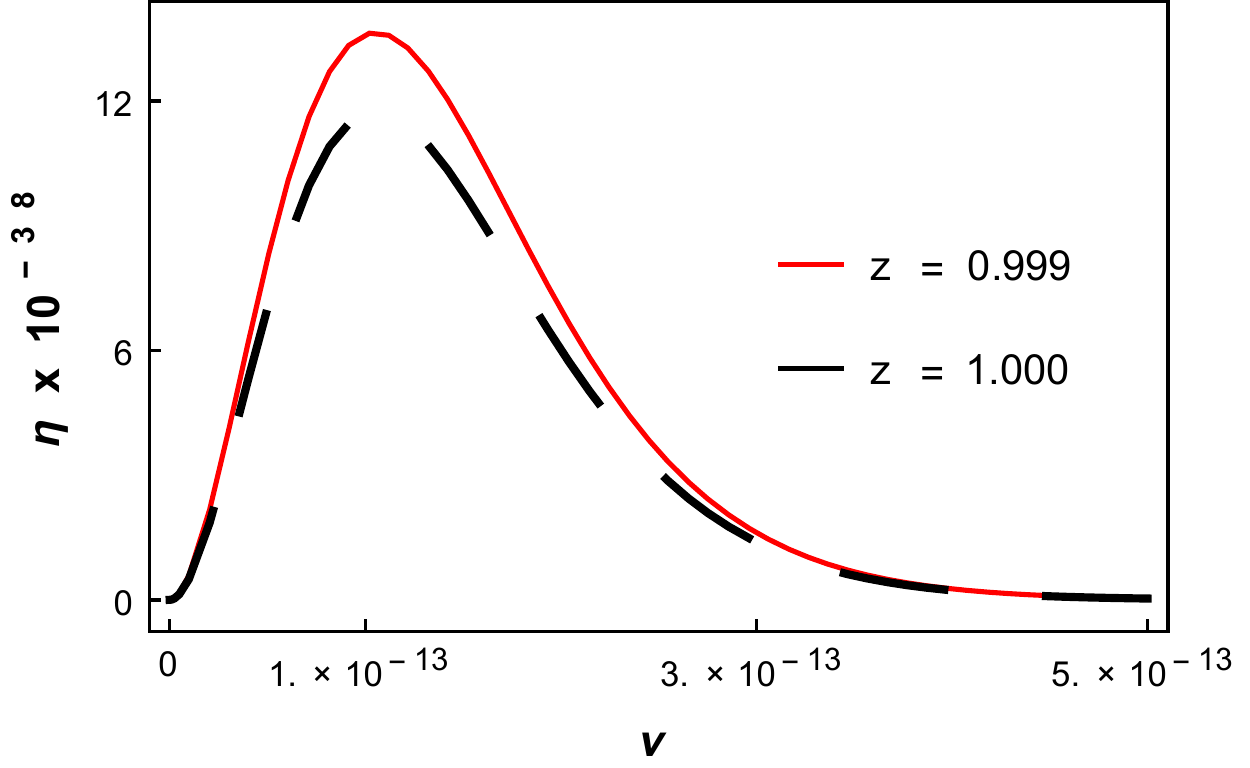} \hspace{0.2 cm} \includegraphics[width=0.4\columnwidth]{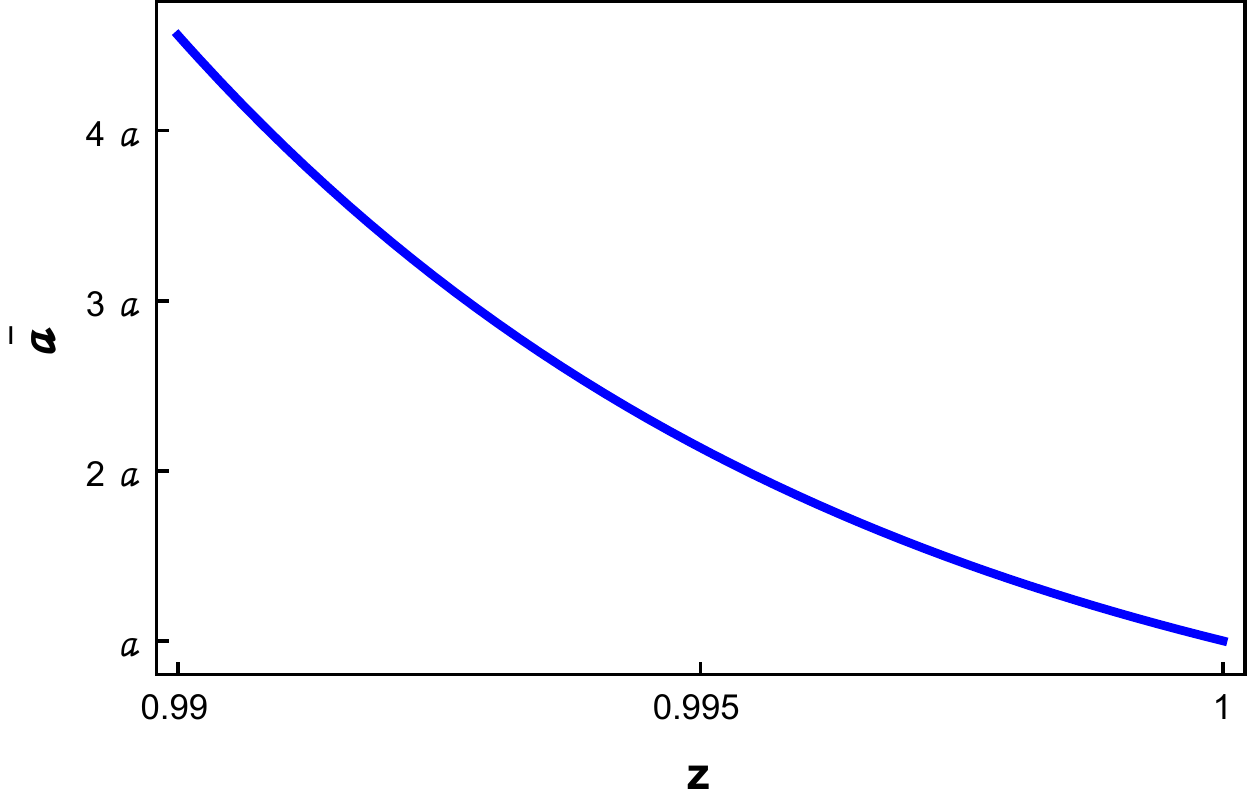}
 \caption{In the left panel we observe the black body radiation curves as a function of the frequency $\nu$ at the CMB temperature for $z\leq 1$. There we can see how values of $z$ smaller than one increase the spectral radiance. In the right panel we show the evolution of the radiation constant $\bar{a}$ as function of $z$, also at the CMB temperature.}
 \label{FIG4_5}
\end{figure}

\begin{figure}[h!]
 \centering
 \includegraphics[width=0.4\columnwidth]{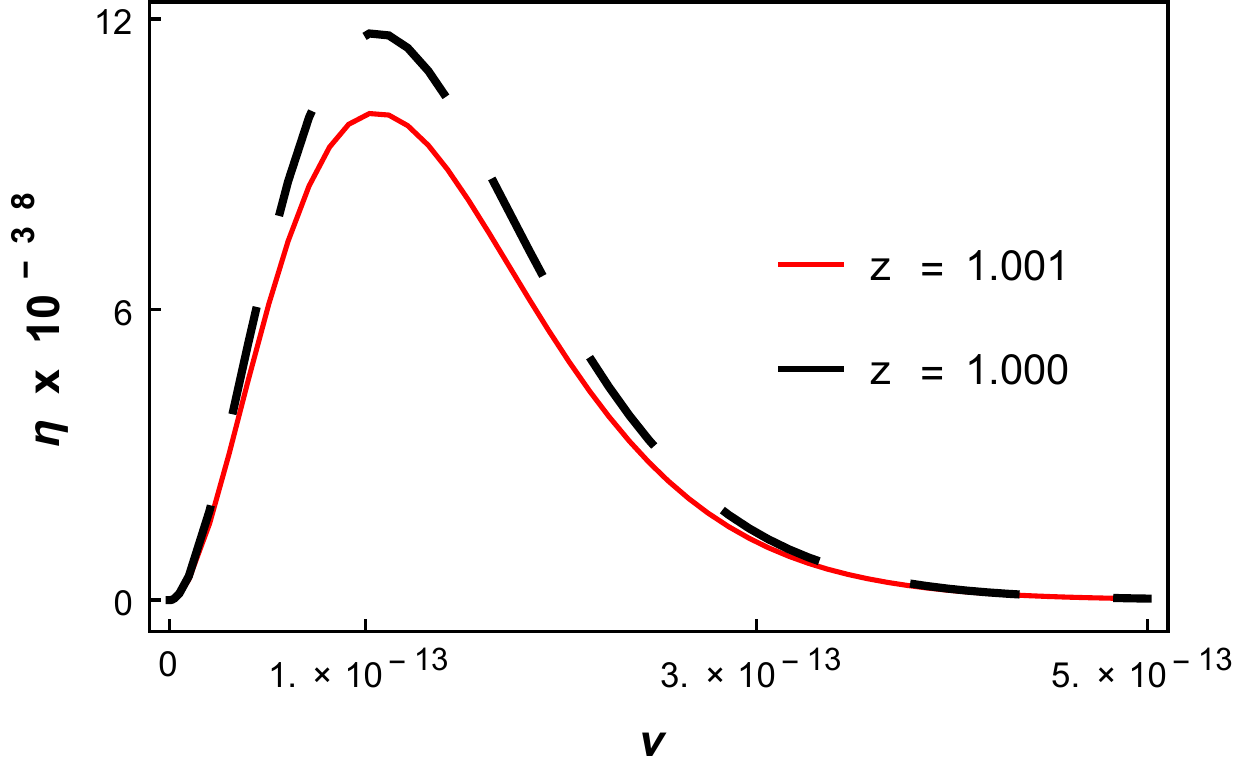} \hspace{0.2 cm} \includegraphics[width=0.4\columnwidth]{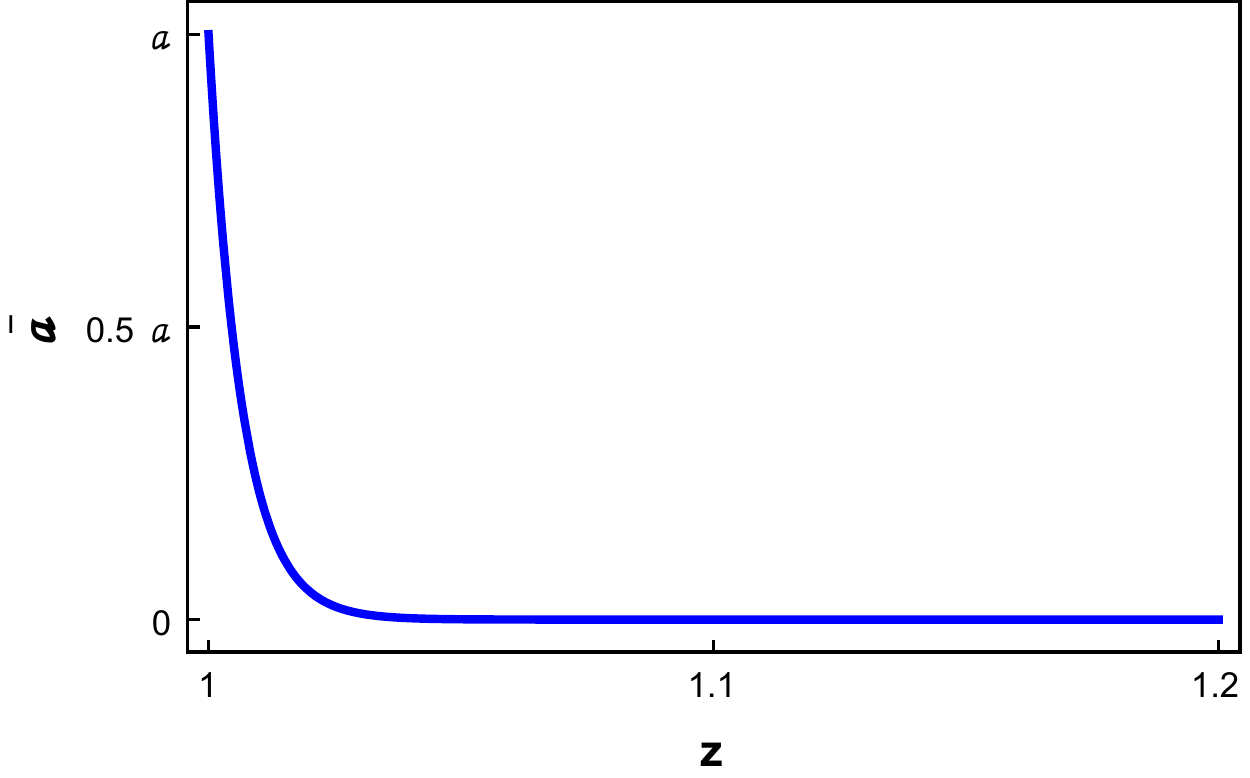}
 \caption{In the left panel we observe the black body radiation curves as a function of the frequency $\nu$ at the CMB temperature for $z\geq 1$. There we can see how values of $z$ greater than one decrease the spectral radiance. In the right panel we show the evolution of the radiation constant $\bar{a}$ as function of $z$, also at the CMB temperature.}
 \label{FIG5}
\end{figure}

{
\subsection{Modified Dispersion Relation}
The Lifshitz-scaling to modified dispersion relation is given by \cite{epassos_16}  
\bea\label{HL01}
\big(E^{2} - \Lambda_{HL}^{-2(z-1)}k^{2z} \big)^{2} - 4 \big(\bar{\xi} R_{HLP}\big)^{2} \big(\Lambda_{HL}\big)^{-2z} E^{4} k^{2z}=0
\eea
%with the parameter $n_{\mu}$ being purely time-like, and 
where $z$ is the Horava-Lifshitz critical exponent. Here $R_{HLP}=\frac{\Lambda_{HL}}{M_P}$ with $M_P=1.22\,\times\,10^{\,19}\,GeV$, and $\Lambda_{HL}$ is the Horava-Lifshitz energy scale. As pointed in \cite{epassos_16}, this ratio can be considered as $R_{HLP} \approx 10^{\,-9}$, unveiling a new route for explicitly breaking supersymmetry without reintroducing fine-tuning \cite{epassos_16,pospelov_12,pospelov_14}. Therefore, the previous constraint lead us to investigate effects of the Horava-Lifshitz crossover scale at the order $\Lambda_{HL} \approx 10^{10}\,GeV$, which follows the restrictions mentioned in \cite{pospelov_12}. 

Following analogous steps applied to the Myers-Pospelov operator, we can now rewrite the dispersion relation (\ref{HL01}) as 
\bea\label{HL02}
\bold{k}_{\rm HL}^{2}= \Lambda_{HL}^{-2(z-1)}E^{2z}\Big( 1 + 2  \big(\bar{\xi} R_{HLP}\big) \big(\Lambda_{HL}\big)^{-z} E^{z} \Big)^{-1}
\eea
Furthermore, one can easily show that taking the Lifshitz critical exponent at $z=1$, we recover Eq.~(\ref{m7}), at $\lambda=-1$.

%%%%%%%%%%%%%%%%%%%%%%%%%%%%%%%%%%%%%%%%%%%%%%%%%%%%%%%%%%%%%%%%%%%%%%%%%%%%%%%%%%%%%
\subsection{Lifshitz-effects on thermodynamic quantities}
To study the effects of Lifshitz critical exponent from the partition function, let us first consider the dispersion relation (\ref{HL02}) to find
\bea\label{HL03}
d\,\bold{k}_{HL}= \left(\frac{z\, E^{z-1} \Lambda_{HL} ^{1-z}}{\sqrt{1+2 \bar{\xi}  R_{HLP}\, E^z \Lambda_{HL} ^{-z}}}-\frac{z\,\bar{\xi}\, R_{HLP}\,E^{2 z-1} \Lambda_{HL} ^{1-2 z}}{\left(1+2\, \bar{\xi} \, R_{HLP}\, E^z \Lambda_{HL} ^{-z}\right)^{3/2}}\right)\,dE.
\eea
Thus, Eqs.~(\ref{HL02})--(\ref{HL03}) can be used to compute the number of available states of the system --- see Eq.~(\ref{T01}) assuming the replacements $\bold{k}^{2} \to \bold{k}_{HL}^{2}\; {{\rm and}\; d\bold{k}^{2} \to d\bold{k}_{HL}^{2}}$. Such an approach allows us to derive the modified partition function
\bea\label{HL04}
\ln {\cal Z}(\beta, V)_{HL} =-\frac{z}{\pi^{2}}\,V\int_{0}^{\infty}\,\frac{E^{3 z-1} \Lambda_{HL} ^{3-z} \left(\bar{\xi}\,R_{HLP}\,E^z+\Lambda_{HL} ^z\right) \sqrt{1+2\, \bar{\xi}\, R_{HLP}\, E^z \Lambda_{HL} ^{-z}}}{\left(2\, \bar{\xi}\,  R_{HLP} \,E^z+\Lambda_{HL} ^z\right)^3}\,\ln\big[1 - e^{-\beta E}\big]\, dE\,.
\eea

There is also a relationship between the Lorentz symmetry breaking parameter and the Lifshitz critical exponent  \cite{epassos_16} that reads
\be
\bar{\xi}(z)=\frac{1.02\,\times\, 10^{\,28\, z-41}}{2.9^{\,2\, z}-0.81^{\,2\, z}}\,.
\ee
%as an anisotropic upper-limit controlled by $z$. Such a value is going to be used to analyze the spectral radiance features, as well as the behavior of the radiation constant. 
The evolution of $\bar{\xi}$ with respect to $z$ is shown in Fig. \ref{FIG4}, where we can see that the Lorentz-violating parameter quickly increases as $z$ gets bigger. By repeating the procedures of section \ref{SBL}, we find that the internal energy, and spectral radiance are explicitly given by
\be\label{enHL}
U_{\,HL}=-\left(\frac{\pa \ln {\cal Z}(\beta, V)}{\pa \beta}\right)_{V}\,,
\ee
and
\be
\eta=\frac{z\, E^{3 z} \Lambda_{HL} ^{3-z} \left(\bar{\xi}\,R_{HLP}\, E^z+\Lambda_{HL} ^z\right) \sqrt{1+2\, \bar{\xi}\, R_{HLP}\, E^z \,\Lambda_{HL} ^{-z}}}{\pi ^2 \left(e^{\beta  \,E}-1\right) \left(2\, \bar{\xi}\,  R_{HLP}\, E^z+\Lambda_{HL} ^z\right)^3}\,;\qquad E = h\,\nu\,.
\ee
Moreover, the energy density follows from Eq. $(\ref{enHL})$ as in the form
\bea
u_{HL}(\beta, z) = \frac{U_{HL}}{V}= u_{_{1\,HL}}(\beta, z) + u_{_{2\,HL}}(\beta, z) \,,
\eea
%which is the internal energy density for the Myers-Pospelov black body with Horava-Lifshitz effect. 
where
\begin{subequations}
\bea
u_{_{1\,HL}}(\beta, z)= \frac{z\, E^{3 z} \Lambda_{HL} ^{3-z} \,\Lambda_{HL} ^z\, \sqrt{1+2\, \bar{\xi}\, R_{HLP}\, E^z \,\Lambda_{HL} ^{-z}}}{\pi ^2 \left(e^{\beta  \,E}-1\right) \left(2\, \bar{\xi}\,  R_{HLP}\, E^z+\Lambda_{HL} ^z\right)^3}\,, 
\eea
\bea
u_{_{2\,HL}}(\beta, z)= \bar{\xi}\,\frac{z\, E^{4 z}\, \Lambda_{HL} ^{3-z} \,R_{HLP}\, \sqrt{1+2\, \bar{\xi}\, R_{HLP}\, E^z \,\Lambda_{HL} ^{-z}}}{\pi ^2 \left(e^{\beta  \,E}-1\right) \left(2\, \bar{\xi}\,  R_{HLP}\, E^z+\Lambda_{HL} ^z\right)^3}\,. 
\eea
\end{subequations}
Thus, in order to find the influence of the Lifshitz critical exponent to the radiation constant, we need to analyze the behavior of
\be 
\bar{a}\equiv u_{HL}\,\beta^{\,4}= \left( u_{_{1\,HL}}(\beta, z) + u_{_{2\,HL}}(\beta, z)\right)\,\beta^{\,4}.
\ee
The panels of Figs. \ref{FIG4_5} and \ref{FIG5} unveil how the Lifshitz critical exponent changes the form of the spectral radiance and the values of the radiation constant at the CMB temperature. It is remarkable that small variations of $z$ lead to effects which could be perceived at low values of temperature such as the in the CMB. Besides, as a matter of consistence, for $z=1$ we recover the standard radiation constant.

}

%%%%%%%%%%%%%%%%%%%%%%%%%%%%%%%%%%%%%%%%%%%%%%%%%%%%%%%%%%%%%%%%%%%%%%%%%%%%%%%%%%%%%%%%
%%%%%%%%%%%%%%%%%%%%%%%%%% - FIGURES - %%%%%%%%%%%%%%%%%%%%%%%%%%%%%%%%%%%%%%%%%%%%%%%%%

%%%%%%%%%%%%%%%%%%%%%%%%%%%%%%%%%%%%%%%%%%%%%%%%%%%%%%%%%%%%%%%%%%%%%%%%%%%%%%%%%%%%%%%%

%%%%%%%%%%%%%%%%%%%%%%%%%%%%%%%%%%%%%%%%%%%%%%%%%%%%%%%%%%%%%%%%%%%%%%%%%%%%%%%%%%%%%%%%

%%%%%%%%%%%%%%%%%%%%%%%%%%%%%%%%%%
\section{Conclusions}
\label{conclu}

In this paper we focused on thermal dynamics from the theories with higher derivative Lorentz-violating operators of quantum electrodynamics extended with dimension-five operators proposed by Myers and Pospelov. 
The Lorentz-violating dimension-five operator predicts a modified dispersion relation, which can be characterized in a covariant form. Such configuration corresponds to a subset of Lorentz invariance violating operators which preserves the rotational invariance and one natural choice for the preferred frame is the frame of the Cosmic Microwave Background (CMB).
 By using the deformed black body radiation and deformed Stefan-Boltzmann law in Myers-Pospelov Lorentz-violating theory, we have shown that the dimension-five operators for the photon sector becomes more effective in very high energies (or temperatures). As expected, such operators are more suitable to probe the physics of extremely high energies where the quantum gravity takes place. By varying the Lorentz-violating parameter, one can see that as it flows from weak to strong strength it probes the UV scale. This is a behavior expected to happen according to the recently developed theory of quantum gravity well-known as Horava-Lifshitz gravity. We can see that the Lorentz-breaking parameter is truly effective only if we deal with high temperatures for the Universe, eg., as the inflationary epoch temperature. We make numerical integrations to evaluate the radiation constant for a Lorentz-breaking parameter variating from $10^{-32}$ to $10^{-13}$ to conclude that when the Universe was really hot, such as slightly before its primordial inflationary phase, the radiation constant was really different from its actual value. Another interesting feature of our modified radiation constant, emerges when we compare it with the one found in a Maxwell Lagrangian coupled to the Carroll-Field-Jackiw (CFJ) term. We conclude that in such approach, by using the CMB temperature we can see that the CFJ Lorentz-violating parameter is more sensible to this scale of temperature than the Myers-Pospelov Lorentz-violating parameter. This behavior is based on the fact that the Myers-Pospelov Lagrangian is written in terms of a dimension-five operator, which is higher than the one coming from the Maxwell-Carroll-Field-Jackiw Lagrangian. On the other hand, by considering a dispersion relation from the Horava-Lifshitz theory which scales with the Lifshitz critical exponent $z$ one can flow from higher to lower dimensional operators, where $z=1$ corresponds to dimension-five operators. Interesting enough, our results corroborate that the operator at $z<1$ develops stronger radiation constant than operator at $z>1$ at CMB temperature. However fingerprints of such operators can be seen in the deformed black body radiation when $z$ is around unit. 

%%%%%%%%%%%%%%%%%%%%%%%%%%%%%%%%%%%%%%%%%%%%%%%%%%%%%%%%%%%%%%%%%%%%%%%%

%\label{conc}

%%%%%%%%%%%%%%%%%%%%%%%%%%%%%%%%%%%%%%%%%%%%%%%%%%%

%\section{Appendix: 3D-Momentum Integral in the Euclidean-space}
%\section{Appendix A: 4D- in the Minkowski-space}

%%%%%%%%%%%%%%%%%%%%%%%%%%%%%%%%%%%%%%%

{\acknowledgments} 

We would like to thank CNPq and CAPES for partial financial support. AM also acknowledges financial support from Universidade Federal de Pernambuco Edital Qualis A.

%\begin{thebibliography}{50}

\end{document}